\documentstyle[osa,manuscript,psfig]{revtex}

\begin{document}

\titlepage
\title{Hyperon polarization in $\nu_\mu$ charged current interaction
at the NOMAD energies}
\author {Liang Zuo-tang$^{1,3}$, and 
Liu Chun-xiu$^{1,2}$}
\address{$^1$Department of Physics,
Shandong University, Jinan, Shandong 250100, China\\
$^2$Institute of High Energy Physics, The Chinese Academy of Sciences,
Beijing 100039, China\\
$^3$The Abdus Salam International Center for Theoretical Physics, 
34100 Trieste, Italy}

\maketitle

\begin{abstract}
We show that, in $\nu_\mu N\to \mu^-\Lambda X$ 
at the NOMAD energies, 
it is impossible to separate 
the products of the fragmentation of the struck quark 
from those of the nucleon remnant.
The latter has a large contribution 
even in the current fragmentation region 
and has to be taken into account in calculating 
$\Lambda$ polarization using different pictures.   
Based on this, we make a rough estimation for 
the  longitudinal $\Lambda$ polarization  
in  $\nu_{\mu} N \to \mu^- \Lambda X$ at the NOMAD energies.
A comparison with the data is given and predictions for 
$\nu_\mu N\to\mu^-\Sigma^+X$ are presented. 
\end{abstract}

%\narrowtext
%\twocolumn
\newpage

Longitudinal hyperon polarizations in lepton induced reactions 
have attracted much attention recently 
(See, e.g., [\ref{Jaffe91}-\ref{Man2001}] 
and the references given there)
since they 
provide an important tool to study the spin transfer in high 
energy fragmentation processes. 
Spin transfer in fragmentation 
is defined as the probability 
for the polarization of the fragmenting quark 
to be transferred to the produced hadron. 
It is one of the important issues in the spin effects 
in the fragmentation processes. 
The study is of particular interest because 
it is still unclear whether the SU(6) or the 
DIS picture is suitable for the description 
of this problem. 
Hyperon polarization in lepton induced reactions 
is an ideal place to study this problem since here the
polarization of the fragmenting quark 
can easily be calculated and 
the hyperon polarization can 
easily be measured by measuring 
the angular distribution of its decay products. 
At sufficiently high energies 
the hadrons produced in the current fragmentation region 
of the deeply inelastic lepton-nucleon scattering 
can be regarded as being completely from  
the struck quark fragmentation. 
Therefore it can be used to study the spin
transfer from the fragmenting quark to the produced hadron. 
Theoretical calculations using different models 
have been made and predictions 
have been given\cite{Kotz98,Ma0002,LXL2001,Ansel01}.

It is encouraging to see that 
measurements of $\Lambda$ polarization in 
deeply inelastic scattering have been carried out by 
HERMES at HERA\cite{HERMES} 
and NOMAD at CERN\cite{NOMAD}. 
They respectively measured the $\Lambda$ polarization 
in $e^+p\to e^+ \Lambda X$ at $E_e=27.6$ GeV 
and in $\nu_\mu N\to \mu^-\Lambda X$ at, 
on the average, $E_\nu=44$ GeV. 
In particular, 
NOMAD has measured it with rather high accuracy 
and has obtained the $P_\Lambda$'s for both 
target and current fragmentation regions, 
hence also its $x_F$ dependence. 
[Here, $x_F\equiv 2p_\parallel /W$, 
$p_\parallel$ and $W$ are respectively 
the momentum component of the produced hyperon 
parallel to the virtual boson direction 
and the total energy of the produced hadronic system 
in the center of mass (c.m.) frame of the system]. 
However, if we compare the NOMAD data\cite{NOMAD} with the 
theoretical predictions in 
[\ref{Kotz98},\ref{Ma0002}-\ref{Ansel01}], 
we see the following distinct difference: 
While all the theoretical results go to zero 
when $x_F$ goes to zero, the data show that 
$|P_\Lambda|$ rises monotonically when $x_F$ 
decreases from positive to negative.
It does not go to zero when $x_F$ goes to zero. 
What does this tell us?
Does it imply that none of the pictures 
in [\ref{Kotz98},\ref{Ma0002}-\ref{Ansel01}]
for spin transfer in high-energy fragmentation processes 
is applicable in $\nu_\mu N\to \mu^-\Lambda X$?

In this note, we show that 
the answer to the question should be ``No!''
This is because, 
in all of the above-mentioned 
calculations\cite{Kotz98,Ma0002,LXL2001,Ansel01}, 
only the struck quark fragmentation has been taken into account 
and the influence from the fragmentation of the remnant of 
the scattered nucleon was neglected. 
This is a good approximation only at high energies. 
In contrast, in the NOMAD or HERMES energy region,  
the influence of the fragmentation of the 
nucleon remnant is usually very large. 
We have to take them into account in particular 
in the region near $x_F=0$. 
We will show our results from a Monte-Carlo 
calculation using the event generator {\sc lepto}\cite{lepto} 
and make a rough estimation of the $\Lambda$ polarization 
by taking the fragmentation of 
the nucleon remnant into account    
using a valence quark model. 

We recall that, 
in deeply inelastic lepton-nucleon scattering, 
we are usually envisaged with the following picture: 
During the collision, 
a quark (or an anti-quark) was struck by the 
exchanged virtual boson.
Viewed in the c.m. frame of the hadronic system, 
the struck quark (anti-quark) flies 
in the opposite direction as the nucleon remnant. 
The system, which consists of 
the struck quark and the nucleon remnant, 
then fragments into the produced hadrons. 
At sufficiently high $W$, 
the hadronization products which contain the struck quark 
and those which contain one or more of the valence quarks 
of the nucleon remnant are usually the leading 
particles in the two opposite directions. 
Their $x_F$'s are usually large in magnitude  
and opposite in sign  
so that they are well separated from each other. 
When we study the hyperon polarization 
in the large and positive $x_F$ region, 
i.e. in the current fragmentation region, 
we need only to take the fragmentation of the struck 
quark into account 
but the influence of the nucleon remnant can be neglected. 
However, in e.g. the NOMAD experiments, 
where $\nu_\mu N\to \mu^-\Lambda X$ is studied, 
the incident energies of $\nu_\mu$  
lies between $10$ and $50$ GeV,  
and $W$ is only several GeV. 
In this case, 
the number of the produced hadrons is usually very small. 
The $|x_F|$ for the hadrons which contain 
the struck quark or one or more of the valence quarks 
of the nucleon remnant can also be very small. 
It is even possible that they move in the opposite 
direction as the quark did before the fragmentation.
As a consequence, it is impossible to 
separate the fragmentation products of the 
struck quark and those of the nucleon remnant 
from each other, even at relatively large $x_F$. 
In particular, in the region of $x_F\sim 0$, 
the contributions from the nucleon remnant can be very important. 

To study this effect explicitly, 
we now take $\nu_\mu p\to \mu^-\Lambda X$ as an example. 
At the relatively low energies, 
we consider only the valence quarks of proton 
thus are envisaged with the following picture: 
In the collision, a $d$-quark in proton is 
knocked out and converted to a $u$-quark, 
which flies in the opposite direction 
as the remaining $(uu)$-diquark. 
This $u$-$(uu)$ system then fragments into hadrons.
There are following possibilities to produce a $\Lambda$ 
which contains one of the three $u$-quarks:
\begin{itemize}
\item[(1)] directly produced and contains the struck $u$; 
\item[(2)] directly produced and contains one of the $u$ 
from the $(uu)$-diquark; 
\item[(3)] decay product of a directly produced hyperon 
resonance that contains the struck $u$; 
\item[(4)] decay product of a directly produced hyperon 
resonance that contains the two $u$'s or 
one of them from the $(uu)$-diquark.  
\end{itemize}
In the cases (1) and (3), 
the vacuum excitation of 
a diquark-anti-diquark pair is needed. 
There should be at least 
one more baryon and an anti-baryon produced,
i.e., there should be 
at least two baryons and one anti-baryon 
produced in the reaction.
In the cases (2) and (4), 
there is no need for 
such vacuum excitation
and there can be only one baryon produced. 
Since the total c.m. energy 
of the $u$-$(uu)$-system $W$ is only several GeV,  
the probabilities for the cases (1) and (3) 
should be much more suppressed compared with those 
for the cases (2) and (4). 
Also because the energy of any one of 
the $u$'s in the $(uu)$-diquark is quite low, 
the $\Lambda$ produced in case (2) and (4) may not 
necessarily fly in the same direction as the original 
$u$-quark. It can also be in the opposite direction. 
In particular in the $x_F\sim 0$ region, 
the contributions from the cases (2) and (4) can dominate.

Since whether 
the cases (2) and (4) dominate and, 
if yes, how strongly they dominate the $\Lambda$ production 
in $\nu_\mu p\to \mu^-\Lambda X$,  
are questions independent of the spin transfer in fragmentation process, 
we can study them using 
a hadronization model which give a good description of the 
properties of the hadrons produced in unpolarized reactions. 
We thus use the event generator {\sc lepto}\cite{lepto} 
based on Lund model\cite{Lund}
to make the numerical estimations.
We generated about $10^7$ $\nu_\mu p\to\mu^-\Lambda X$ events, 
analyzed the origins of the produced $\Lambda$'s  
and obtained the results shown in Fig.1.
From the figure, we see clearly that the contributions 
from the cases (2) and (4) are important. 
They are much higher than those from the cases (1) and (3) 
in the region near $x_F=0$. 
Hence, to calculate the polarization 
of $\Lambda$ in $\nu_\mu p\to\mu^-\Lambda X$ at such 
energies, in particular for $x_F\sim 0$, 
we have to take the fragmentation of the 
nucleon remnant into account.

Having seen that the contributions from 
the nucleon remnant are indeed very important 
in the NOMAD energy region, 
we now calculate the $\Lambda$ 
polarization in $\nu_\mu N\to\mu^-\Lambda X$ 
by taking them into account. 
Apparently, the detailed results should 
depend very much on the polarizations of the quarks in 
the nucleon remnant, which are unclear yet. 
It is thus impossible 
to make a detailed calculation at present. 
However, since the energies are relatively low, 
the characteristic features of the results 
should be mainly determined by the valence quarks.
We therefore make a rough estimation 
using a valence quark model to calculate the 
polarizations of the quarks in the nucleon remnant 
in the following.

Since we take only the valence quarks 
in the nucleon into account, 
the leading order hard subprocess 
in $\nu_\mu N \to \mu^- HX$
can only be $\nu_\mu d \to {\mu}^- u$. 
According to the standard model for electro-weak interaction,
the polarization of the 
outgoing struck quark is $P_u^{(stru)}=-1$.
We neglect the mass of the quarks, hence 
the helicity of the quark is conserved 
in the process $\nu_\mu d \to {\mu}^- u$.
So the $d$-quark was also in the 
helicity ``$-$'' state before the scattering. 
For the reaction $\nu_\mu p \to \mu^- HX$, 
the nucleon remnant is a $(uu)$-diquark. 
If the proton was in the helicity ``+'' state, 
the remaining $(uu)$-diquark has to be 
in $(uu)_{1,-1}$ state. 
(Here we use the subscripts 
to denote the spin and its component 
along the moving direction 
of the outgoing struck quark 
in the c.m. frame of the hadronic system. 
The minus sign is because the moving direction of 
the outgoing struck quark is opposite to that of the diquark.) 
The relative probability, obtained from 
the SU(6) wave function, is $2/3$.
If the proton is in the helicity ``$-$'' state,  
the remaining $(uu)$-diquark has to be 
in $(uu)_{1,0}$ state, 
and the relative probability is $1/3$. 
Hence, the $u$-quark in the nucleon remnant 
is polarized and the polarization is $P_u^{(di,p)}=-2/3$.
For the reaction $\nu_\mu n \to \mu^- HX$, 
the nucleon remnant is a $(ud)$-diquark. 
Similarly, we obtain that 
this $(ud)$-diquark can be 
in the $(ud)_{1,-1}$, 
$(ud)_{1,0}$ or $(ud)_{0,0}$ state, 
and the relative probability is $1/6$, 
$1/12$ or $3/4$ respectively. 
Hence, the polarization of the $u$- or 
$d$-quark in the neutron remnant 
is $P_u^{(di,n)}=P_d^{(di,n)}=-2/3$ 
if it is from the spin-1 $(ud)$-diquark or 
$P_u^{(di,n)}=P_d^{(di,n)}=0$ 
if it is from the spin-0 $(ud)$-diquark.

All these polarizations can be transferred to the 
hyperons produced in the hadronization. 
We calculate the polarizations 
for the hyperons of the different origins 
in the following ways:
For those which are directly produced 
and contain the struck quark 
or one of the quarks in the nucleon remnant, 
we calculate in the same way 
as we did in [\ref{BL98},\ref{LL2000},\ref{LXL2001},\ref{XLL2002}]
for $q\to HX$ using two different pictures, 
i.e. the SU(6) and the DIS pictures. 
For those which are directly produced 
and contain the diquark,  
we calculate using the SU(6) wave-function. 
The polarization of the $\Lambda$ in the final state in 
$\nu_\mu p\to\mu^-\Lambda X$ is given by,
\begin{equation}
P_{\Lambda}=[P_u^{(stru)}t^F_{\Lambda,u}\langle n_\Lambda^{(1)}\rangle+
P_u^{(di,p)}t^F_{\Lambda,u}\langle n_\Lambda^{(2)}\rangle+
\sum_jt^D_{\Lambda,Hj}P_{Hj}\langle n_{\Lambda,Hj}\rangle]/ 
\langle n_\Lambda^{(total)}\rangle.
\end{equation}
where $\langle n^{(1)}_{\Lambda}\rangle$ 
and $\langle n^{(2)}_{\Lambda}\rangle$ 
are the average numbers of 
the $\Lambda$'s of the origins 
(1) and (2) mentioned above respectively, 
and $\langle n_{\Lambda,H_j}\rangle$ is that 
from the decay of $H_j$'s;
$P_{H_j}$ is the polarization of $H_j$;
$\langle n^{(total)}_{\Lambda}\rangle$ 
is the total number of $\Lambda$'s.
These average numbers are determined 
by the hadronization mechanisms, and we just 
use the results obtained from {\sc lepto}. 
$t^F_{\Lambda,u}$ 
is the fragmentation polarization transfer factor, 
which is the probability for 
the polarization of $u$-quark to be transferred 
to $\Lambda$ if the $\Lambda$ contains this $u$-quark.
It is different in the SU(6) or the DIS picture. 
$t^D_{H_i,H_j}$ is the decay polarization transfer factor 
determined by the decay process. 
For details of $t^F_{\Lambda,u}$ and $t^D_{H_i, H_j}$, 
see e.g. [\ref{XLL2002}] and the references given there. 

After the calculations,  
we obtain the longitudinal polarization of $\Lambda$ 
in $\nu_\mu p\to\mu^-\Lambda X$ and 
$\nu_\mu n\to\mu^-\Lambda X$ at $E_\nu=44$GeV 
as shown in Figs.2a and 2b.
To show the influence of the nucleon remnant fragmentation, 
in the figures, we show also the results 
in the case when only the 
struck quark fragmentation is taken into account. 
From Fig.2a,  
we see that the contributions from 
the $(uu)$-diquark fragmentation 
are indeed very large, even at $x_F>0$.
In fact, 
the characteristic features of the results 
in the region around $x_F\sim 0$ 
are mainly determined by the $(uu)$-diquark fragmentation.
E.g., we see that, in this region, 
$P_{\Lambda(L)}$ is negative 
and $|P_{\Lambda(L)}|$ increases with decreasing $x_F$.
This is determined by the decay contribution 
of the $\Sigma^{*+}$'s 
which contain the $(uu)$ diquarks. 
It contributes negatively to $P_{\Lambda(L)}$ 
and the contribution increases with the decreasing $x_F$. 
Compare the results in Fig.2a with those in Fig.2b, 
we see that, 
$P_{\Lambda(L)}$ for $\nu_\mu n\to\mu^-\Lambda X$
differs greatly from that for $\nu_\mu p\to\mu^-\Lambda X$.
It is very small in the whole $x_F$ region. 
This is because the contributions from the decay of 
the heavier hyperons which contain the $(ud)_1$
or a $u$- or a $d$-quark from $(ud)_1$ are different 
in sign and cancel with each other.
Since NOMAD has approximately an isobar target, 
the NOMAD data\cite{NOMAD} corresponds approximately 
to a mixture of  $(2/3)$ n- and $(1/3)$ p-targets. 
We add the results with these weights 
and compare the obtained results with the data\cite{NOMAD} in Fig.2c.
We see that the qualitative features of the results 
agree with the data\cite{NOMAD}.

We also made similar calculations for 
$P_{\Sigma^+(L)}$ in $\nu_\mu N \to \mu^- \Sigma^+ X$ and
obtained the results in Fig.3. 
The advantage to study $\Sigma^+$ instead of $\Lambda$ is 
that the decay contributions to $\Sigma^+$ 
are much smaller. 
The major contribution to $\nu_\mu p\to\mu^-\Sigma^+X$
is from the $(uu)$ fragmentation. 
The $\Sigma^+$'s which contain the $(uu)$-diquark 
dominate for most $x_F$. 
This is why we see that the 
obtained $P_{\Sigma^+(L)}$ is negative 
and its absolute value is very large 
for $\nu_\mu p \to \mu^- \Sigma^+ X$. 
These can be checked by future experiments.

It should be noted that, 
almost at the same time, another group  
has carried out\cite{EKN2002} the similar calculations 
as presented in this paper independently 
using a different model for the spin transfer 
in the fragmentation of diquarks. 
The aim of the authors was to use 
$\Lambda$ polarization in such process to 
study the polarization of the strange sea in nucleon. 
Assuming that there is a correlation between the 
polarization of the strange sea and 
that of the struck quark 
(which was not included in our calculations), 
they obtained a good fit 
to the NOMAD data\cite{NOMAD}. 
The authors also 
found similar effect which confirms the 
conclusion that we present above and 
briefly reported earlier in the conference talks\cite{LXL2001}. 
We also note that, in our calculations, we used 
the default set of parameters in {\sc jetset}, and such 
set of parameters may not, 
as pointed out in [\ref{EKN2002}], 
reproduce the data on the average numbers of the 
strange baryons produced in the process. 
However, since we are interested in the 
relative contributions of $\Lambda$ from the 
different sources 
(1) to (4) mentioned above, this may be not very 
sensitive to our results in particular the 
qualitative conclusions we reached above. 
More precisely, there should be little influence 
from the parameters such as 
the overall strange suppression etc. 
But there can be some influence from the parameters 
such as decuplet baryon production suppression etc
since a large part of the contribution from the 
diquark fragmentation come from $\Sigma^*$ decay. 
To see whether this is true, we tuned 
the related parameters and repeated the calculations. 
We found out that 
there are indeed some influence on the 
quantitative results 
but the qualitative features remain.

In summary, 
we show that the products from the   
fragmentation of the struck quark and those of the nucleon remnant 
can not be separated from each other in the NOMAD energy region.
The latter gives the dominant contribution in 
the region around $x_F\sim 0$ 
and the characteristic features of the 
hyperon polarization in this region 
are determined by this contribution.
We made a rough estimation of the $\Lambda$ polarizations 
by taking the fragmentation of the nucleon remnant into account 
and the results are consistent with the NOMAD data. 
Further predictions for 
$\nu_{\mu} N \to \mu^- \Sigma^+ X$ are made.

We are in debt to Dr. D. Naumov, who suggested us to 
make the research in the target fragmentation region 
and continuously communicated with us 
on different aspects. 
This work was supported in part by the National Science Foundation
of China (NSFC), the Education Ministry of China 
under Huo Ying-dong Foundation,
and the Postdoctoral Science Foundation of China.

\noindent

\vskip 0.2cm

%\newpage

\begin{figure}[ht]
\psfig{file=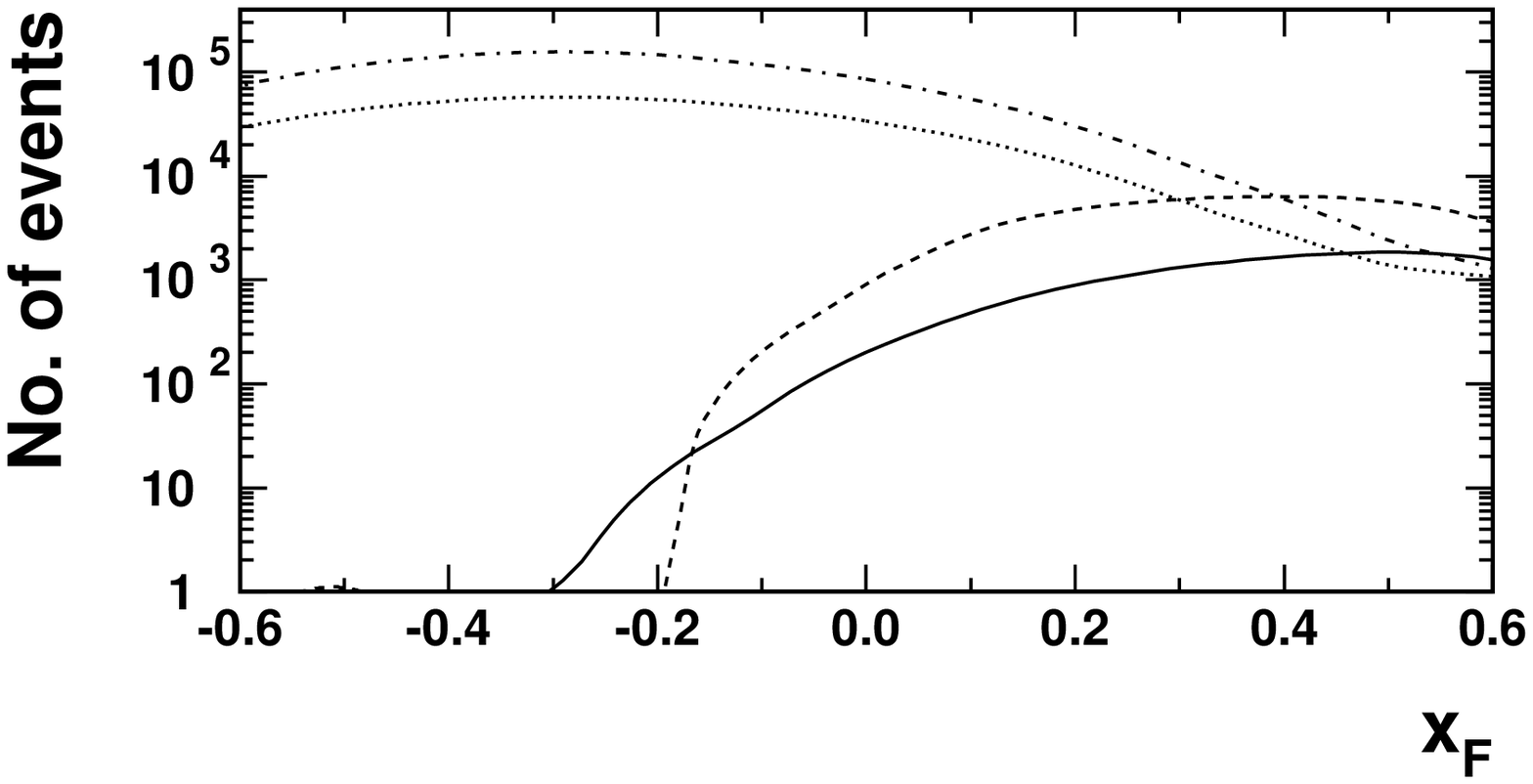,width=6.8cm} 
\caption{Different contributions 
to $\Lambda$ in $\nu_\mu p\to \mu^-\Lambda X$ at $E_\nu=44$GeV. 
Here, the solid, dotted, dashed and dash-dotted lines 
denote respectively 
(1) directly produced and contain the struck quark;
(2) directly produced and contain a $u$-quark in $(uu)_1$; 
(3) decay products of hyperons which contain the struck quark; 
(4) decay products of hyperons 
    which contain the $(uu)_1$ or a $u$ in the $(uu)_1$. } 
\label{fig1}
\end{figure}

\begin{figure}[ht]
\psfig{file=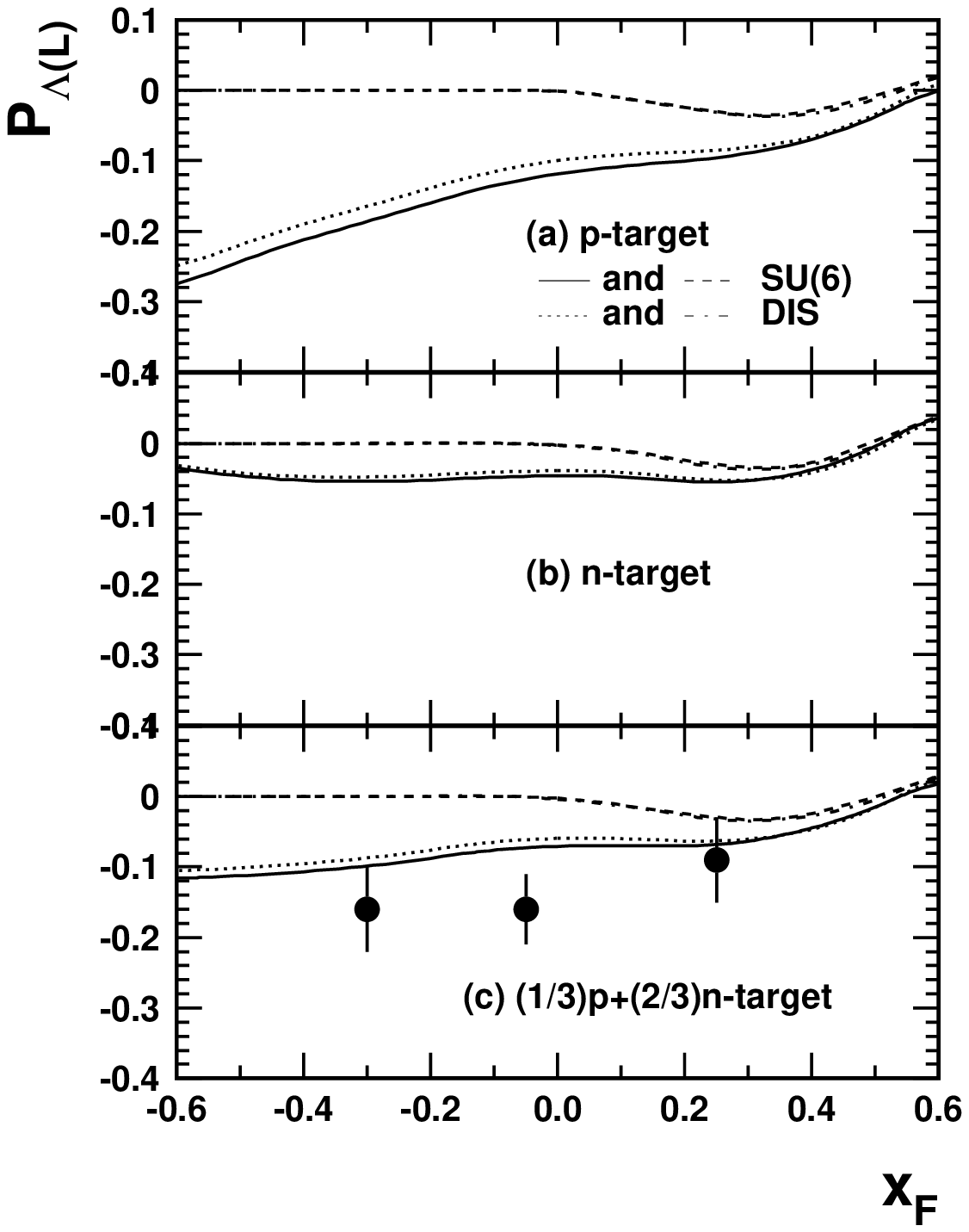,width=8cm}
\caption{ $P_{\Lambda(L)}$ 
as a function of $x_F$ in $\nu_\mu N\to
\mu^-\Lambda X$ at $E_\nu=44$ GeV 
when both the contributions from the fragmentation of 
the struck quark and that of the nucleon remnant are 
taken into account (solid and dotted lines),
compared to the results when only the struck quark fragmentation 
is taken into account 
(dashed and dash-dotted lines). 
The data are taken from [\ref{NOMAD}].}
\label{fig2}
\end{figure}

\begin{figure}[ht]
\psfig{file=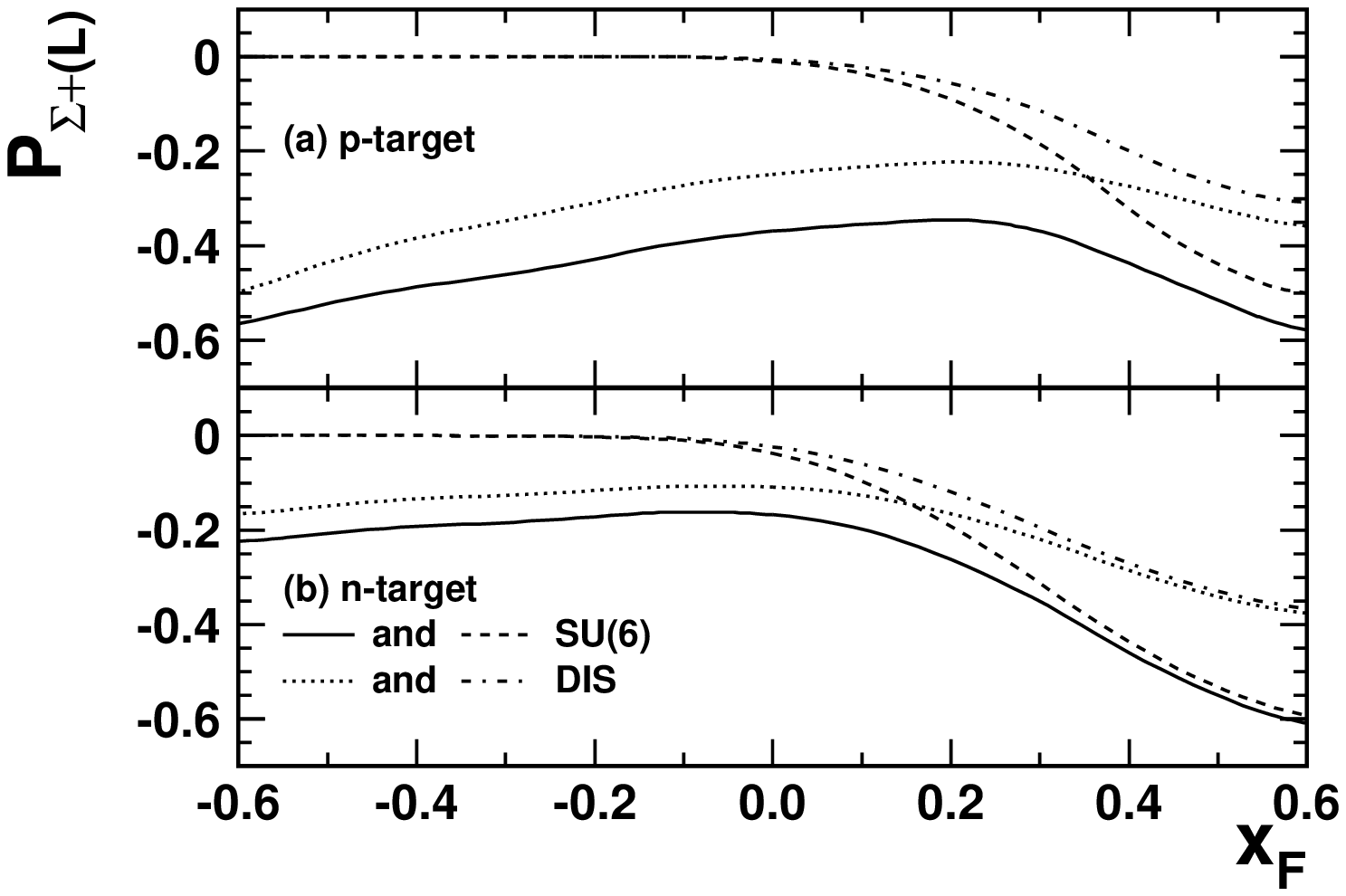,width=6.8cm}
\caption{$P_{\Sigma^+(L)}$ 
as a function of $x_F$ in $\nu_\mu N\to
\mu^-\Sigma^+ X$ at $E_\nu=44$ GeV 
when both the contributions from the fragmentation of 
the struck quark and that of the nucleon remnant are 
taken into account (solid and dotted lines),
compared to the results when only the struck quark fragmentation 
is taken into account 
(dashed and dash-dotted lines). }
\label{fig3}
\end{figure}


\begin {thebibliography}{99}
\bibitem{Jaffe91} R.L. Jaffe, and X. Ji,
          Phys. Rev. Lett. {\bf 67 }, 552 (1991);
          Nucl. Phys. B{\bf 375}, 527 (1992);
          M. Burkardt and R.L. Jaffe,
          Phys. Rev. Lett. {\bf 70}, 2537 (1993);
          R.L. Jaffe, Phys. Rev. D{\bf 54}, R6581 (1996).
\label{Jaffe91}
\bibitem{GH93} G.Gustafson, J.H\"akkinen,
               Phys.Lett.B{\bf 303},350(1993).
\label{GH93}
\bibitem{BL98} C. Boros, and Z. Liang, 
             Phys. Rev. D{\bf 57}, 4491 (1998).
\label{BL98}
\bibitem{Kotz98} A. Kotzinian, A. Bravar, D. von Harrach,
                Eur. Phys. J. C{\bf 2}, 329 (1998).
\label{Kotz98}
\bibitem{AL99} D. Ashery, H.J. Lipkin, Phys. Lett. B{\bf 469}, 263 (1999);
         hep-ph/0002144.
\label{AL99}
\bibitem{Ma2000} B.Q. Ma, I. Schmidt, and J.J. Yang,
         Phys. Rev. D{\bf 61}, 034017 (2000).
\label{Ma2000}
\bibitem{ABM00} M.Anselmino, M.Boglione and F.Murgia,
 Phys. Lett. B{\bf 481}, 253 (2000).
\label{ABM00}
\bibitem{LL2000} C. Liu and Z. Liang, Phys. Rev.
               D{\bf 62}, 094001 (2000).
\label{LL2000}
\bibitem{Ma0002} B.Q. Ma, I. Schmidt, J. Soffer, and J.J. Yang,
         Phys. Rev. D{\bf 62}, 114009 (2000).
\label{Ma0002}
\bibitem{LXL2001} C. Liu, Q. Xu, and Z. Liang,
    Phys. Rev. D{\bf 64}, 073004 (2001); 
    Z. Liang, in Proc. of the 31st Inter. Sympo. on 
    Multiparticle Dynamics, Datong, 2001, 
    World Scientific (2002), p.78, hep-ph/0111403; 
    and Proc. 3rd Circum-Pan-Pacifc Sympo. on High Energy Spin Phys., 
    Beijing, 2001, hep-ph/0205017;
    C. Liu, Q. Xu, Z. Liang, {\it ibid}.
\label{LXL2001}
\bibitem{Ansel01} M.Anselmino, M.Boglione, U.D'Alesio, and F.Murgia,
Eur. Phys. J. C{\bf 21}, 501 (2001).
\label{Ansel01}
\bibitem{XLL2002} Xu Qing-hua, Liu Chun-xiu and Liang Zuo-tang, 
                  Phys. Rev. D{\bf 65}, 114008 (2002).
\label{XLL2002}
\bibitem{HERMES} HERMES Collaboration, A. Airapetian {\it et al}., 
Phys. Rev. D{\bf 64}, 112005 (2001).
\label{HERMES}
\bibitem{NOMAD} NOMAD Collaboration, P. Astier {\it et al}.,
   Nucl. Phys. B{\bf 588}, 3 (2000); {\bf 605}, 3 (2001);
   D. Naumov, in Proc. of the 14th Inter. Sympo. on Spin Phys., 
   AIP Conf. Proc. 570 (2001), p.489, hep-ph/0101325.
\label{NOMAD}
\bibitem{Man2001} M.L. Mangano {\it et al}., hep-ph/0105155.
\label{Man2001}
\bibitem{lepto} G Ingelman, A.Edin, J.Rathsman, 
        Comp. Phys. Comm. {\bf 101}, 108 (1997).
\label{lepto}
\bibitem{Lund} B.~Anderson, {\it et al.}, Phys. Rep. {\bf 97}, 31 (1983).
\label{Lund}
\bibitem{EKN2002} J.Ellis, A. Kotznian, and D. Naumov, hep-ph/0204206.
\label{EKN2002}

\end{thebibliography}
\end{document}